# Exploring the Suitability of QUIC for the Internet of Things

**Carles Gomez, Nika Soltani-Tehrani,** *Universitat Politècnica de Catalunya, Castelldefels, 08860, Spain*

**Jon Crowcroft,** *University of Cambridge, Cambridge, CB3 0FD, United Kingdom*

***Abstract—QUIC is an emerging transport-layer protocol that provides reliability and security. QUIC was designed to overcome issues from other protocol stacks used in the Internet, such as TCP/TLS, especially focusing on web traffic performance improvement. Therefore, QUIC was not conceived for Internet of Things (IoT) scenarios, which are characterized by significant resource constraints. However, as QUIC's prominance increases, and the IoT continues to expand, QUIC may offer connectivity opportunities for IoT devices. In this paper, we explore the suitability of QUIC for IoT environments. Leveraging optional functionality, we propose, discuss, and evaluate a QUIC profile for IoT scenarios that is currently being considered for IETF standardization.***

The World Wide Web (WWW) has been the main driver for the massive Internet expansion that started in the early 90s and has connected billions of users. The HyperText Transfer Protocol (HTTP) has been a primary enabler of the WWW, allowing clients to interact with web servers on top of the Transmission Control Protocol (TCP). The latter has been the main transport-layer protocol providing reliability in the Internet since the Flag Day, in January 1983. Subsequently, Transport Layer Security (TLS) was inserted between HTTP and TCP in order to secure web communication.

The HTTP/TLS/TCP protocol stack has carried a large fraction of Internet traffic. However, the Internet evolution, and the aim for performance improvement, have shown several limitations of the two lower layers of this stack, TLS and TCP [1]. Firstly, these two protocols require each an initial handshake between the two involved endpoints (to establish security parameters, and to initialize a connection, respectively), before upper-layer messages can be sent. Therefore, each protocol's handshake introduces one Round Trip Time (RTT) before application data can actually be transmitted. Secondly, a TCP connection may suffer a Head of Line (HoL) blocking problem, whereby the loss of a segment will force subsequent segments to wait in the sender's buffer until the lost one has been recovered, even if the latter segments carry upper-layer content that is not part of the one carried by the recovered segment (e.g., different web page objects). Thirdly, a TCP connection is bound to the IP addresses (and ports) of the involved endpoints. However, a network change (e.g., from a 5G connection to a Wi-Fi network) typically leads to an IP address change, which terminates any existing TCP connection. Finally, middleboxes (e.g., firewalls, NAT routers, and load balancers) can inspect TCP headers and interfere with the expected TCP behavior, as well as reject unrecognized new formats or header field values, hindering protocol evolution.

As a result, more than one decade ago, Google engineers started the design of QUIC, a transport-layer protocol intended to overcome the aforementioned issues of TCP and TLS. Rather than being a clean-slate transport-layer protocol, QUIC partly integrates modern functionality of both TCP and TLS in order to provide reliability and security, merging some of their features for efficient operation.

The main QUIC specifications were released in 2021

[2], following significant deployment experience. In parallel, HTTP/3, the latest HTTP version as of the writing, defined QUIC as its underlying transport-layer protocol [3]. Furthermore, an increasing number of applications are being provided with QUIC support, such as the Domain Name System (DNS), media, tunnels, and telemetry, among others. In consequence, QUIC adoption on the Internet is growing significantly, and it has been considered that QUIC has potential to replace TCP as main transport-layer protocol on the Internet [4].

As QUIC establishes itself as a dominant protocol, its support may become crucial to allow device connectivity. However, QUIC was not designed considering the features of important scenarios that were emerging in parallel, such as Internet of Things (IoT) environments. Many IoT scenarios are based on constrained-node networks, where simple devices (e.g., equipped with sensors and/or actuators) exhibit significant computational and energy constraints, and their networks often offer low bit rates and a relatively high error rate (since many IoT technologies are wireless). On the other hand, the IoT is playing a fundamental role in our society, as it is enabling smart use cases in a wide range of application domains, including smart homes, smart cities, smart industry and smart health. Estimates indicate that there are ~20 billion of connected IoT devices as of the writing, and such numbers are projected to continue increasing, representing the main driver for the Internet growth nowadays [5].

In this paper, we explore the suitability of QUIC for IoT environments. Leveraging optional functionality, we propose, discuss and evaluate a QUIC profile for IoT scenarios that is currently being considered for IETF standardization [6]. As a benchmark, we compare QUIC's performance and characteristics with those of the transport-layer functionality embedded in the Constrained Application Protocol (CoAP) [7]. The latter was developed on purpose for constrained-node network environments. Despite some limitations, a properly configured QUIC can offer a reasonably good performance in many IoT scenarios and use cases.

## RELATED WORK

Several research works have investigated the use of QUIC in an IoT context [8-12]. However, most of such works focus on joint application-layer and transport-layer performance evaluation, instead of exploring how QUIC can be adapted for improved operation in IoT scenarios.

Kumar et al., and Fernández et al., evaluated the performance of Message Queuing Telemetry Transport (MQTT) on top of QUIC. They found that the latter outperforms TCP as underlying protocol for MQTT in some environments, while offering results similar to TCP's in others [8, 9]. Iqbal et al. reached similar conclusions regarding QUIC and TCP as underlying protocols for the Advanced Message Queuing Protocol (AMQP) [10]. Herrero analyzed the performance of CoAP on top of UDP, TCP and QUIC, concluding that QUIC offers a lower message loss probability [11].

Instead of considering application- and transport-layer performance, Eggert evaluated QUIC implementations on two different IoT platforms, in terms of transfer time, memory usage and energy consumption [12]. Among others, that work illustrated approaches to reduce the memory footprint of QUIC implementations. However, it did not explore an optimized QUIC protocol configuration for IoT scenarios.

## QUIC OVERVIEW

QUIC provides end-to-end reliability by means of features mostly inherited from TCP, including advanced developments of the latter. QUIC uses selective acknowledgments (ACKs) and retransmits lost data, either by means of a Fast Retransmit mechanism similar to TCP's or, in some cases, after the expiration of a Probe TimeOut (PTO). While the baseline congestion controller specified in QUIC is based on TCP's NewReno, a sender can opt to use a different congestion controller.

QUIC also integrates TLS 1.3 to provide security, using an initial handhsake that jointly negotiates transport and security parameters. As a result, QUIC reduces the connection establishment delay before communication by one RTT, compared with separate TCP and TLS operation. For the first connection between a client and a server, the initial handshake adds one RTT before upper-layer data can be transmitted via so-called 1-RTT packets. For subsequent connections between the same client and server, it is possible to integrate the handshake with data transmission by using so-called 0-RTT packets.

QUIC sends packets that may contain one or more frames. The latter may carry upper-layer data (e.g., by using STREAM frames) or signaling information. A QUIC packet may include frames of different streams. Lost data recovery applies only to the affected streams, avoiding TCP's HoL problem. Several QUIC packet header fields are protected, to minimize middlebox interference. QUIC packets are carried by UDP packets, for easier integration with existing Internet infrastructure.

In QUIC, a connection is independent of the IP addresses and ports of the involved endpoints. A QUIC connection is identified by means of a Connection ID.

# COAP OVERVIEW

CoAP is an application-layer protocol designed for constrained-node networks. Accordingly, it provides lightweight operation and a significant degree of flexibility, along with security. CoAP comprises two sublayers: the upper one handles the interaction between a client and a server by means of requests and responses, following the REST model used also in the WWW; the lower sublayer provides transport-layer functionality, since CoAP was originally designed to run atop UDP. Hereinafter, and for the sake of comparison with QUIC, in this paper we will refer to CoAP focusing on its lower sublayer functionality.

CoAP defines two types of messages that may carry requests and responses: Confirmable (CON) messages and Non-confirmable (NON) messages. The former elicit ACKs from the destination endpoint, whereas the latter do not. The transmission of CON messages follows a stop-and-wait behavior, with timer-based retransmission and exponential back-off for congestion control. A Retransmission TimeOut (RTO) is started for every CON message that is sent.

There exist several alternatives to secure CoAP messages. One relies on Datagram Transport Layer Security (DTLS), which introduces at least two RTTs before communication. It is also possible to use Object Security for Constrained RESTful Environments (OSCORE) to protect CoAP message payloads and a subset of header fields. OSCORE can be configured to avoid initial handshakes by utilizing a pre-shared security context between the involved endpoints.

CoAP also supports group communication, which enables multicast use cases.

# PROFILING QUIC FOR THE IOT

This section presents and discusses a QUIC profile for the IoT, which is based on our related proposal currently being considered for IETF standardization [6]. The QUIC dimensions considered are the following: connection handling, congestion window, reliable/unreliable transmission, header overhead, ACK rate, initial RTT, and multicast.

## Connection handling

An IoT device may use energy-saving techniques, such as radio duty-cycling. However, such techniques require the IoT device to remain in low-energy consumption mode (i.e., *sleep* mode) for relatively long intervals. In order to establish a QUIC connection, an IoT device is recommended to act as client, thus initiating connection establishment. This approach allows the IoT device to send the connection establishment messages when appropriate (i.e., during an active interval).

A typical IoT use case is a device that periodically provides sensor reading reports to another system for several years. Considering the resource constraints of many IoT scenarios, it is important to minimize the protocol overhead in QUIC. To this end, the lifetime of a QUIC connection should ideally be as long as possible. The *max_idle_timeout* parameter in QUIC determines the maximum time during which a QUIC connection can remain inactive (i.e., without any sent or received packets). An infinite connection lifetime can be configured upon agreement of the two involved endpoints at connection establishment.

However, many Internet paths include middleboxes, which may discard UDP datagrams after an inactive period (typically, of 30 seconds) [2]. QUIC offers two options to avoid this problem. The first one requires one of the two endpoints to send a packet containing a PING frame as keep-alive message before the middlebox inactivity timer expiration. The second option relies on opening a new connection with 0-RTT transmission between the same client and server pair when the client (i.e., the IoT device) has data ready to transmit, after the middlebox inactivity timer has expired. The 0-RTT connection establishment involves the transmission of 2 Initial QUIC packets from the client and 1 such packet from the server. For security and path capacity verification reasons, the UDP packets carrying Initial packets (one of them coalesced with the 0-RTT packet carrying the user data) are padded up to a UDP payload size of at least 1200 bytes (a size of 1200 bytes is assumed hereinafter). Despite this overhead, there are conditions whereby the 0-RTT reconnection approach can outperform the keep-alive approach. Assuming an IoT device that sends a sensor reading periodically every $T$ seconds, there exists a threshold value $T_{Thr}$. For $T < T_{Thr}$, the first approach (QUIC keep-alive), offers lower overhead. Otherwise, the second approach (0-RTT reconnection) is more efficient. This is illustrated in Figure 1. For an upper-layer data payload of 10 bytes, $T_{Thr}$=42.5 minutes. Note that, in some use cases, $T$ may be in the order of hours or even days [13].

The same issue can also be addressed from the application layer, by means of heartbeat messages sent at a sufficiently high rate, to avoid expiration of the middlebox inactivity timer. An advantage of this approach is that it allows liveness checks at the application level. However, it incurs additional overhead.

## Congestion Window

Appropriate initial and minimum congestion window sizes need to be configured for some types of memory-constrained IoT devices. Class 1 IoT devices are equipped with ~10 kB of RAM [14]. While devices of this class are common in IoT environments, they cannot accommodate the initial congestion window that should be used per the QUIC specification [15], typically equal to ten times the maximum datagram size, which is about ~15 kB. On the other hand, in contrast with TCP, where the minimum congestion window is one segment, the minimum congestion window value recommended in QUIC is two packets (of maximum datagram size). Some IoT devices that run TCP are only able to support a window size of one Maximum Segment Size (MSS) [16], and therefore cannot exploit the greater minimum congestion window in QUIC.

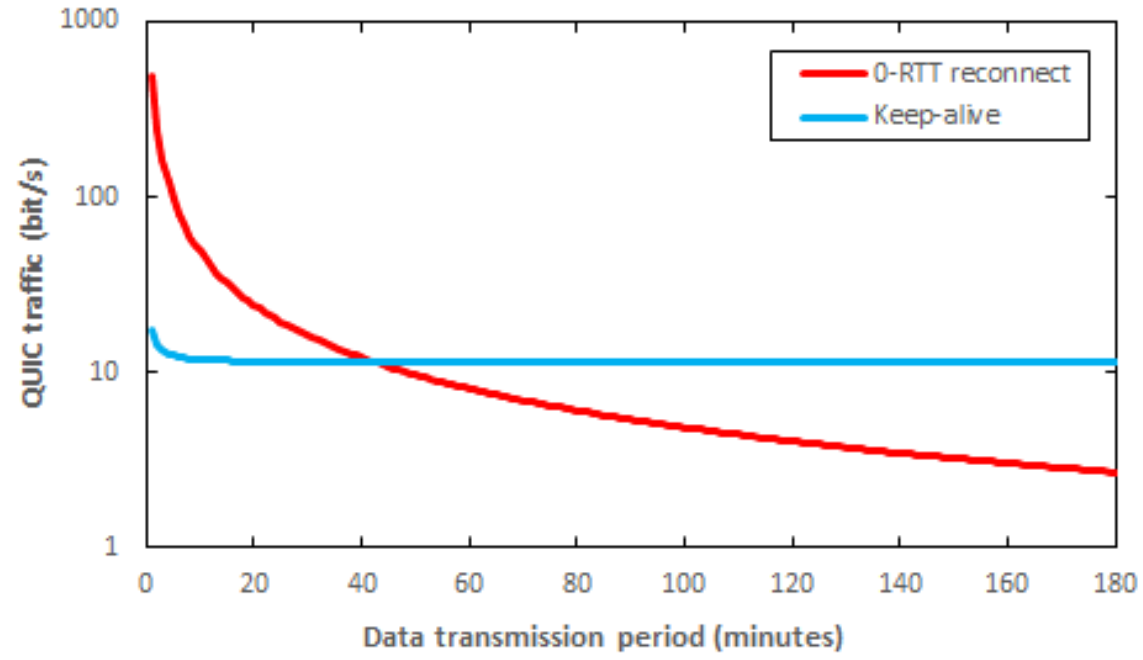

FIGURE 1. QUIC connection establishment/maintenance traffic in steady state, as a function of the data transmission period, *T*, for the keep-alive and the 0-RTT reconnect approaches. A middlebox UDP inactive period (and keep-alive period) of 30 seconds; the shortest, IoT-optimized possible QUIC header overhead (see Figure 2); and a 10-byte user data payload have been assumed.

## Reliable versus Unreliable Transmission

IoT applications present a diversity of requirements in terms of reliable delivery. Some use cases tolerate a certain degree of packet loss (e.g., a sensor that frequently takes and transmits a temperature reading for offline analysis), whereas reliable delivery is of the utmost importance in others (e.g., a safety system that needs to send an alarm message upon detection of a fire). QUIC was designed to offer reliability by default, by means of Acknowledgments (ACKs) from the receiver and retransmission of lost packets. Subsequently, DATAGRAM frames were defined [17]. While such frames are ACK-eliciting, they are never retransmitted. Therefore, in QUIC, IoT data that does not require reliable delivery can be carried by DATAGRAM frames. This allows to reduce protocol overhead, since packet loss is common in many IoT scenarios.

## Header Overhead

Due to the energy constraints of many IoT devices, it is fundamental to reduce the size and the number of the data units to be transmitted and received. Furthermore, lower time on air allows to reduce collisions, which may be a problem in scenarios with a high spatial device density.

The packet format of QUIC offers flexibility, since fields such as Connection ID and packet number are of variable size. Therefore, it is possible to reduce QUIC packet header overhead, as long as scenario characteristics allow it. However, related trade-offs need to be assessed.

The possible size of a Connection ID field, which is present twice in a 0-RTT packet, and once in a 1-RTT packet, ranges from 0 to 20 bytes. Decreasing Connection ID size negatively affects privacy and load balancing efficiency. However, IoT environments where these issues are tolerable to a certain degree can exploit shorter Connection ID fields.

On the other hand, the size of a packet number, which is present once per packet, can be between 1 and 4 bytes. A shorter packet number size increases packet identification ambiguity. However, many IoT devices will send packets at a very low rate (e.g., one sensor reading per minute, per hour or even per day). In such case, or if the ACK rate is sufficiently high, a short packet number can be safely used.

Note that, in many cases, a single stream per connection will sufice for a simple IoT device that transmits sensor readings and receives configuration commands. Therefore, in such cases, a packet will carry a single STREAM frame carrying data.

## ACK Rate

As described earlier, many IoT use cases are based on devices sending a single data packet infrequently. Despite the minimum congestion window of two packets recommended in QUIC, in such use cases there will be at most a single data packet in flight. If the receiver does not generate an application-layer response, it will wait before sending the ACK for up to 25 ms by default. This behavior follows the Delayed ACKs mechanism originally defined for TCP [18], which aims to decrease the number of ACKs or packets transmitted by the receiver. However, in the described single-packet use case, delaying the ACK needlessly contributes delay and may increase energy consumption of the sender IoT device, without decreasing the packet overhead.

A QUIC extension currently being developed offers a sender two ways to request a receiver to send an ACK immediately upon receipt of a packet. First, the sender may transmit an IMMEDIATE_ACK frame along with the STREAM frame carrying the application data, which increases the packet size by one byte. In this case, only the next ACK will be sent without an additional delay. Second, the sender may also send an ACK_FREQUENCY frame requesting the maximum ACK delay to be zero, so that ACKs sent by the receiver are not delayed thereafter. However, such behavior will only be possible if the receiver also supports it.

For more capable IoT devices in scenarios that allow a large congestion window size to be used, requesting an ACK rate lower than default (i.e., one ACK every 2 received packets) allows to reduce the number of ACKs transmitted by the receiver, decreasing energy consumption and bandwidth usage of the involved devices. Such request can be carried out by sending an ACK_FREQUENCY frame including an appropriately requested ACK rate.

## Initial RTT

QUIC maintains an RTT estimate for various purposes, including loss detection, congestion control and connection liveness. The initial value for the RTT estimate is equal to 333 ms by default. The RTT is one input parameter in the algorithm that calculates the Probe TimeOut (PTO). For the default intial RTT, the default initial PTO value is 1 second. On the other hand, the PTO value is also used as input to other parameters, such as *max_idle_timeout*, which is greater than 3·PTO, and the persistent congestion duration, with a recommended value of 3·PTO.

The default value for the initial RTT in QUIC is not suitable for many IoT scenarios, where the RTT may be greater. Several possible reasons include the following: i) low physical-layer bit rates, which in some cases may be in the order of ~100 bit/s, leading to high transmission time (in the order of seconds); ii) radio duty-cycling, which may introduce long intervals (from seconds to days) where the IoT device sleeps and is therefore unreachable; iii) intermittent connectivity provided by sparse satellite constellations, introducing delays up to the order of hours; iv) use of quasi-bidirectional technologies (e.g., some LPWAN technologies), which allow transmission to an IoT device only after the former has explicitly signalled that it is ready for communication by sending a previous message.

In order to avoid undue packet retransmission in the described environments, which would entail unnecessary consumption of energy and bandwidth resources, the initial RTT needs to be set appropriately, according to the characteristics of the specific usage scenario.

## Multicast

Similarly to TCP, QUIC is a unicast protocol. However, some IoT use cases require multicast support. A prominent example is a lighting application, where after the action of a user, a connected switch sends a command message intended to turn on/off a group of lights in a room or in a street. The efficiency of a multicast protocol is particularly needed in resource-constrained IoT scenarios.

# DISCUSSION AND EVALUATION

In this section, we discuss and evaluate QUIC as a protocol for IoT environments, using CoAP as a benchmark. We provide a feature comparison, an energy performance evaluation, and a memory footprint overview.

## Benchmarking QUIC Features against CoAP

The different design principles of QUIC and CoAP significantly influence their features (see Table 1). Despite the performance improvements that QUIC provides over separate TCP and TLS operation, such as the reduced connection establishment latency, QUIC is essentially intended as a reliable and secure, general-purpose protocol for the Internet. QUIC requires a connection establishment, always elicits ACKs for packets carrying user data, and it does not support multicast. QUIC’s greater data transmission overhead (see Figure 2), recommended initial and minimum congestion windows, recommended initial RTT (and, thus, PTO), and ACK rate handling are in line with general Internet scenarios. In contrast, CoAP does not use connections and does not require a handshake to secure communication, it presumes a single-packet transmission window, has a concise header (even including upper-sublayer fields), allows unacknowledged transmission, does not delay ACKs, uses a default RTO greater than QUIC’s initial PTO, and supports multicast.

Nevertheless, in QUIC, it is possible to configure various parameters and mechanisms for improved operation in IoT environments. If required by the scenario characteristics, the QUIC congestion window size can be set to 1 packet, DATAGRAM frames can be used for unreliable (even if acknowledged) transmission, header size can be reduced (see Figure 2), delayed ACK operation can be controlled by the sender, and the initial RTT (and initial PTO) can be increased appropriately.

TABLE 1. Comparison of QUIC and CoAP features.

| Feature | QUIC | CoAP |
|---|---|---|
| Connection establishment | 1-RTT and 0-RTT | No (Initial handshake not needed without security or with OSCORE) |
| Congestion window | 10 packets (default initial), 2 packets (recomm. minimum) | 1 (default) |
| Unreliable delivery | DATAGRAM frame (confirmed) | NON message (unconfirmed) |
| Delayed ACKs | Yes (default) | No |
| ACK rate requests | Yes | No |
| Initial PTO/RTO | 1 s (default) | [2 s, 3 s] (default) |
| Multicast | No | Yes |

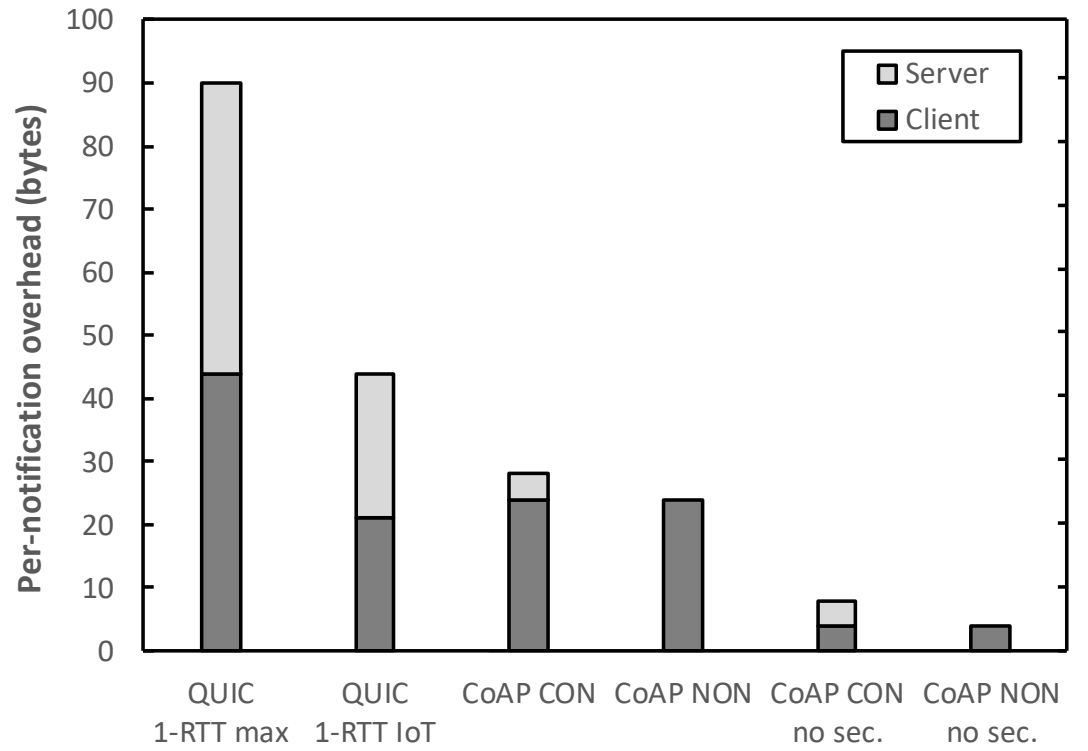

FIGURE 2. Protocol overhead per message containing user data (e.g., a sensor reading), for QUIC and CoAP. QUIC 0-RTT transmission (not plotted) requires 2400 bytes from the client, while 1223 bytes are assumed from the server. For QUIC, maximum and minimum (IoT-optimized) header sizes are considered for 1-RTT packets. A QUIC packet is assumed to carry a single STREAM frame. For CoAP, no optional fields other than OSCORE are added. 16-byte authentication tags are assumed for both QUIC and CoAP. OSCORE is assumed to incur a 20-byte overhead increase. For CoAP, no security is also considered, although it is generally not recommended.

## Energy Performance Evaluation

We next evaluate the lifetime of a battery-operated IoT device that sends a notification (every period $T$), carrying a 10-byte application-layer message using QUIC over UDP, IPv6, 6LoWPAN and acknowledged-mode IEEE 802.15.4 at 2.4 GHz. For the calculations, a coin-cell battery of 230 mAh, the current consumption features of the nRF5340 Nordic device [19], and no packet losses have been assumed. In order to capture a low overhead scenario, the shortest possible headers have been considered for the layers below QUIC, comprising a 7-byte 6LoWPAN-compressed IPv6/UDP header (with global IPv6 addresses), short IEEE 802.15.4 addresses and a star-topology layer-2 network. Two alternatives are considered regarding QUIC connection establishment and maintenance: the Keep-alive approach, where the connection is maintained by using periodic PING frame transmission every 30 seconds (for $T$ > 30 seconds), and the reconnection approach, based on performing a 0-RTT data packet transmission every period $T$. For the Keep-alive approach, the longest and the shortest (IoT-optimized) QUIC packet headers are considered. For comparison, CoAP transmission has also been included in the study, for CON and NON transmission of the same payload size carried with QUIC, with and without Keep-alive mechanisms. QUIC and CoAP ideal (without Keep-alive) approaches are also evaluated, aiming to capture performance in a middlebox-free environment.

As shown in Figure 3, as $T$ increases, sleep intervals become dominant, therefore device lifetime increases. For low values of $T$, QUIC Keep-alive (especially the IoT-optimized one), performs similarly to CoAP CON transmission, while CoAP NON offers the greatest device lifetime. As $T$ increases, QUIC Keep-alive is limited by the overhead of its heartbeats. While the QUIC 0-RTT reconnect approach incurs a high overhead for low $T$, it outperforms QUIC Keep-alive for $T$ greater than ~15 minutes, even approaching ideal results for very large $T$, as it does not require heartbeat traffic.

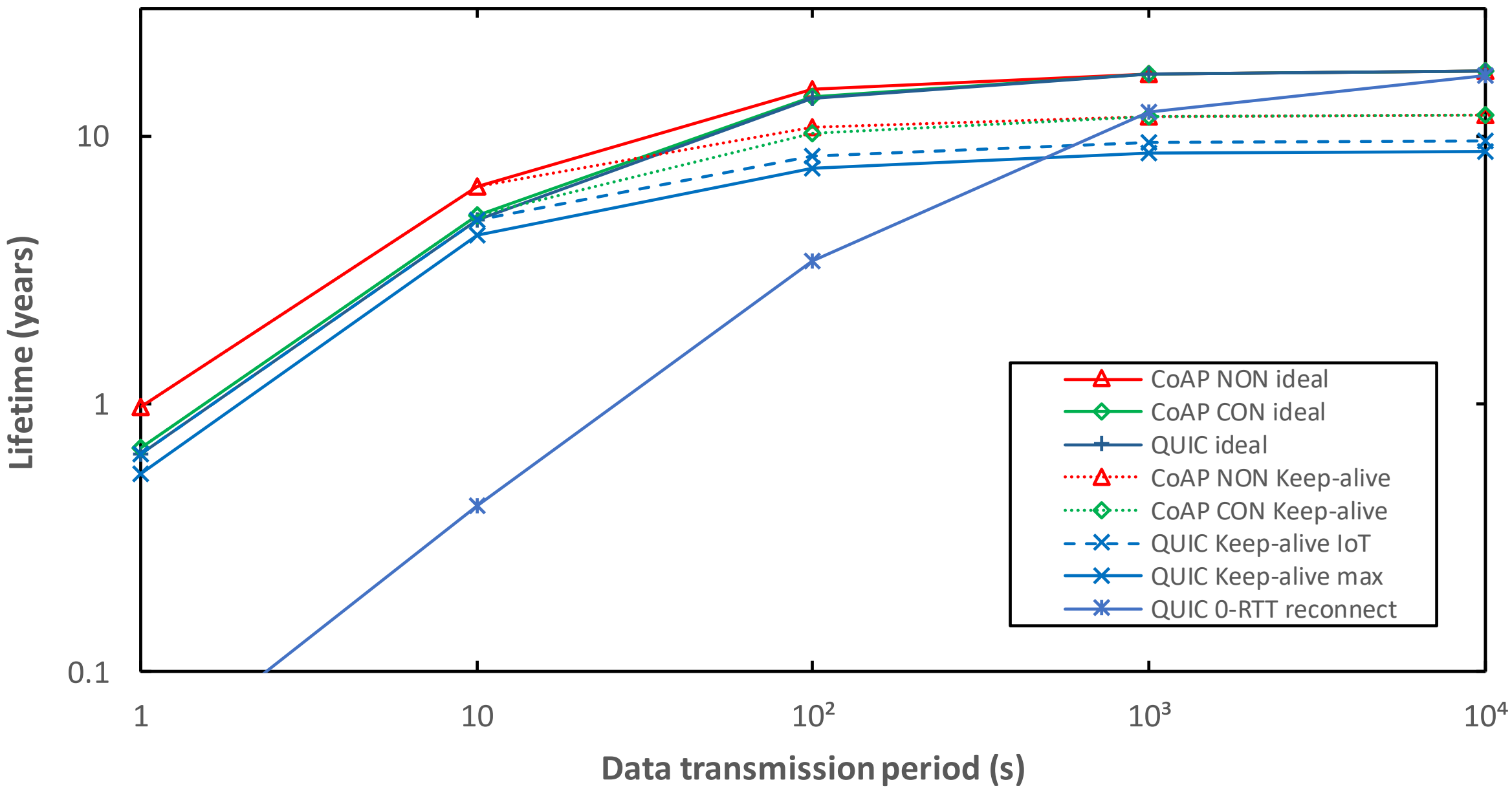

FIGURE 3. Theoretical battery lifetime of an IoT device sending a packet periodically over IEEE 802.15.4, for a battery capacity of 230 mAh, and for various CoAP and QUIC approaches.

Note that, in terms of battery lifetime performance, $T_{Thr}$ is smaller than the one shown in Figure 1. This is due to the hardware and protocol energy overheads associated to packet transmission: such overheads penalize the Keep-alive approach to a greater extent in relative terms, as it involves shorter packets. QUIC cannot generally reach the energy performance of CoAP, due to its greater header and message overhead. Nevertheless, QUIC can achive multiyear battery lifetime even for $T$ in the order of seconds.

## Memory footprint

While there exist CoAP (and TCP [16]) implementations suitable for Class 1 devices, current QUIC implementations consume a higher amount of memory. To our best knowledge, the lowest QUIC memory footprint reported is 58 kB of flash size and 64 kB of heap size [12]. Nevertheless, there are avenues for reducing QUIC memory footprint, such as adjusting send and receive buffer sizes (which in many IoT use cases can be smaller than for general Internet usage), and supporting only the minimum cryptographic functionality.

# CONCLUSION

In this paper, we have explored the suitability of QUIC as a protocol for IoT. We have compared its features with the transport-layer functionality embedded in CoAP. QUIC exhibits limitations when considering IoT scenarios: it requires a connection establishment, data is always acknowledged, and it does not support multicast. However, a properly configured QUIC can offer a reasonably good performance in a wide range of IoT use cases.

# ACKNOWLEDGMENTS

Carles Gomez has been supported in part by the Spanish Government's Ministerio de Ciencia, Innovación y Universidades MCIU/AEI/10.13039/501100011033/FEDER/UE through project PID2023-146378NB-I00, and through the Estancias de Movilidad en Centros Extranjeros de Enseñanza Superior e Investigación PRX24/00397 grant.

**Carles Gomez** received his Ph.D. degree from Universitat Politècnica de Catalunya in 2007. He is a Full Professor at the same university. He is a co-author of numerous technical contributions including papers published in journals and conferences, IETF RFCs, and books. He serves as a co-chair of the IETF 6Lo working group. His research interests focus mainly on IoT and deep-space communications. He is the corresponding author of this article. Contact him at carles.gomez@upc.edu.

**Nika Soltani-Tehrani** received her B.Sc. degrees in Computer Engineering and Electrical Engineering from Amirkabir University of Technology, Iran. She is currently pursuing the Erasmus Mundus Joint Master's Degree in Communication and Data Science (CoDaS) at Universitat Politècnica de Catalunya, and Aalto University. She is a co-author of the IETF Internet Draft providing guidance on the use of QUIC in IoT. Her research interests include IoT, transport protocols, and wireless communications. Contact her at nika.soltani.tehrani@estudiantat.upc.edu.

**Jon Crowcroft** has been the Marconi Professor of Communications Systems in the Computer Laboratory, University of Cambridge since 2001. He has worked in the area of Internet support for multimedia communications for over 40 years. Three main topics of interest have been scalable multicast routing, practical approaches to traffic management, and the design of deployable end-to-end protocols. Contact him at jon.crowcroft@cl.cam.ac.uk.